# Mortality Analysis of Early COVID-19 Cases in the Philippines Based on Observed Demographic and Clinical Characteristics

Roel F. Ceballos

University of Southeastern Philippines, Davao City, Philippines,
https://orcid.org/0000-0001-8267-6482
Email Correspondence: roel.ceballos@usep.edu.ph

*Abstract*

*This study aims to determine the demographic, epidemiologic, and clinical characteristics of COVID-19 cases that are highly susceptible to COVID-19 infection, with longer hospitalization and at higher risk of mortality and to provide insights that may be useful to assess the vaccination priority program and allocate hospital resources. Methods that were used include descriptive statistics, nonparametric analysis, and survival analysis. Results of the study reveal that women are more susceptible to infection while men are at risk of longer hospitalization and higher mortality. Significant risk factors to COVID-19 mortality are older age, male sex, difficulty breathing, and comorbidities like hypertension and diabetes. Patients with these combined symptoms should be considered for admission to the COVID-19 facility for proper management and care. Also, there is a significant delay in the testing and diagnosis of those who died, implying that timeliness in the testing and diagnosis of patients is crucial in patient survival.*

## 1.0 Introduction

In response to the threat of the COVID-19 pandemic, nations have developed and adopted strategies for management, mitigation, and control of the disease transmission in communities, schools, and workplaces. The implementation of community lockdowns and the closure of most establishments were found to be effective in lowering the rate of disease transmission (Ceballos, 2020; Rabajante, 2020) but was also found to have adverse effects on the economy (Philippine Statistics Authority, 2021) and the psychological health of the people (Saladino et al., 2020). Due to the negative effect of lockdowns, it is considered a temporary measure while the vaccination begins. The vaccination program against COVID-19 aims to boost immunity, which is instrumental in promoting social and economic activities. However, there is a challenge in acquiring the supply of vaccines since the demand is high worldwide. To manage effectively the limited supply of vaccines in the Philippines, the Inter-Agency Task Force (IATF), an advisory body responsible for decisions on the management and control of COVID-19 in the Philippines, has rolled out priority groups for the COVID-19 vaccination program to reduce mortality, preserve health system capacity, and protect the populations most-at-risk. (Department of Health, 2021). The decision involving the prioritization framework of vaccination must be supported by statistical evidence to ensure that the aims of the vaccination program are met.

Just like the supply of vaccines for COVID-19, hospital resources are also limited in the Philippines. As of April 2020, there are only around 13.5 hospital beds per 10,000 population in the National Capital Region while only 8.9 beds per 10,000 population in Davao Region. The situation is worst in the Bangsamoro Autonomous Region in Muslim Mindanao (BARMM) with 2.7 beds and MIMAROPA with only one bed per 10,000 population, respectively. According to the World Health Organization, the prescribed proportion of hospital beds is one per 1,000 population (Sanchez, 2021). The total number of hospital beds dedicated to COVID-19 in the Philippines is only 33,584. Out of this total, 3,084 are dedicated Intensive Care Unit (ICU) beds, 18,865 are isolation beds, and the remaining 11,635 are ward beds (Department of Health, 2021).

To maximize hospital resources, the Center for Disease Control (CDC) released the Interim Clinical Guidance for Management of Patients with Confirmed Coronavirus Disease (COVID-19). Hospital guidelines for admission of COVID-19 patients include severity, comorbidity, and symptoms of cases. Severe cases like dyspnea, hypoxia, or more than 50 percent lung involvement on imaging and critical cases like respiratory failure, shock, or multiorgan system dysfunction are highly associated with mortality (Chen et al., 2020; Guan et al., 2020; Huang et al., 2020; Wang, Hu et al., 2020; Wu et al., 2020; Hu et al., 2020). Case fatality is higher for patients with the following comorbidities: cardiovascular disease, diabetes, chronic respiratory disease, hypertension, prior stroke, chronic kidney disease, chronic lung disease, and cancer (Center for Disease Control and Prevention [CDC] COVID-19 Response Team, 2020; Wu et al., 2020; Yang et al., 2020). Age is a vital risk factor for severe infection, complications, and death (Arentz et al., 2020; Guan et al., 2020). Patients with clinical characteristics associated with high case fatality and mortality are usually admitted to hospital facilities for treatment and care.

Aside from mortality, the length of hospital stay is also an important outcome for COVID-19 patients. The importance is highlighted in the availability of hospital beds, ventilators, intensive unit care (ICU) facilities, and doctors during the surge of the disease. Most hospitals have a limited number of ventilators and ICU facilities, and the scarcity is a huge problem when more patients are hospitalized for an extended period. Several studies have reported that 26 percent to 32 percent among hospitalized patients were placed under ICUs and the mortality among these patients was around 72 percent at the most (Wang, He et al., 2020; Wu et al., 2020; Zhou et al., 2020).

This study aims to determine the demographic, epidemiologic, and clinical characteristics of COVID-19 cases that are highly susceptible to COVID-19 infection, with longer hospitalization and at higher risk of mortality and to provide insights that may be useful to assess the vaccination priority program and allocate hospital resources.

**The framework of the Study**
Several variables are identified as important risk factors of subgroup susceptibility, mortality, and longer hospitalization for COVID-19 patients based on several studies (Guan et al., 2020; Wang, Hu et al., 2020; Wu et al., 2020; Zhou et al., 2020). In this study, the researcher analyzed the early records of COVID-19 cases in the Philippines to determine these risk factors. Specifically, the demographic, epidemiologic, and clinical characteristics of patients found in the data from the Department of Health (DOH) drop were used to determine subgroup exposure and susceptibility, mortality, and length of hospital stay.

COVID-19 risk factors that are significantly associated with greater exposure and susceptibility to COVID-19 infection, longer hospital stay, and higher mortality should be included as criteria in determining vaccination priority and the planning and management of medical and hospital resources. The framework in Figure 1 shows the process of utilizing early records of COVID-19 by identifying the risk factors associated with longer hospitalization and mortality to aid as inputs to policy on the management of patients and medical resources and determination of vaccination priority in the population subgroups. For the demographic characteristics, the researcher examined age groups, nature of work related to health care, and sex to determine the most exposed and most susceptible demographic subgroups. For the epidemiologic characteristics, the researcher figured the time intervals between the onset of symptoms to testing (specimen collection) and diagnosis (released of confirmed status) to determine the impact of delay in the testing and diagnosis of COVID-19. Finally, the researcher analyzed which comorbidity and symptoms are associated with a longer hospital stay and higher mortality.

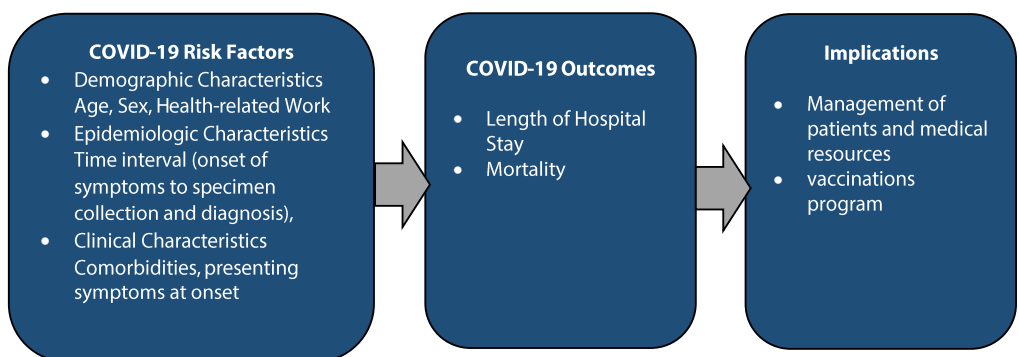

*Figure 1.* Framework of the Study

## 2.0 Methodology
**Research Design and Data**

The study used a retrospective cohort design by utilizing records of confirmed COVID-19 positive cases in the Philippines. The data on demographic, epidemiologic, and clinical characteristics of confirmed COVID-19 patients were retrieved from the case information of the DOH COVID-19 data drop, a data repository provided by DOH for the public. The case information does not contain patient identifiers so the use of these data complies with the Data Privacy Act of 2012. Furthermore, DOH allows the public to access, use, re-use, and share freely the data found in the COVID-19 data drop as stipulated in their guidelines (Department of Health, 2021). A total enumeration of all cases was retrieved from January to April 30, 2020. After that, some variables of patients' case information are no longer available publicly in the DOH data drop. Thus, we were only able to utilize data up to April 30, 2020.

**Data Analysis**

To determine and compare the susceptibility of the different demographic subgroups, descriptive statistics such as frequency and percentage were used. The median and interquartile range were employed to summarize the onset from symptom to testing and diagnosis, length of hospital stay, and age. Nonparametric methods were used in the analysis to compare the length of hospitalization of the patients according to demographic, epidemiologic, and clinical characteristics. Mann-Whitney U test was used to compare dichotomous variables or variables with two categories while the Kruskal-Wallis test was used to compare variables with more than two categories. Chi-square was used to determine the association of demographic, epidemiologic, and clinical characteristics with mortality. The Cox multivariate regression analysis was used to determine demographic, epidemiologic, and clinical risk factors of COVID-19 mortality. For detailed reference on the process of Mann-Whitney U test, Kruskal-Wallis test, and Cox regression analysis, see Zar (1999), Kruskal and Wallis (1952), and Breslow (1975), respectively.

## 3.0 Results

### Susceptibility of the demographic subgroups

One of the aims of this research is to examine the demographic profile of early COVID-19 and identify which subgroups are more susceptible to COVID-19 infection upon exposure. Table 1 shows the number and percentage of cases for the different demographic subgroups during the early stage of the pandemic in the Philippines. The percentage of infected females (54%) is higher by an average of 8 percent than males (46%), which implies that women are more susceptible than men given equal exposure to the virus. In addition, among the total infected females in the population, only 47 or 0.6 percent are pregnant. The result is lower than expected since pregnant women are believed to be more susceptible to COVID-19 because they are vulnerable to respiratory infection (Liu et al., 2020). Furthermore, the most susceptible age group is the adult or working class (71%) followed by the senior citizens (26%). The adult or working class, specifically, the front liners, and essential workers or those whose nature of work requires them to move around constantly (service sector jobs) have greater susceptibility to the COVID-19. On the other hand, the senior citizens are susceptible to infection as they have low immune system and comorbidities. Lastly, the infection rate among health care workers is almost 17 percent. Belingheri, et al. (2021) explain that health workers have greater susceptibility than the general community because the nature of their work during the pandemic requires them to spend long hours in hospitals, leading to lower sleep quality, lower immunity against infection, and greater risks of exposures. According to the results of the systematic review conducted by Bandyopadhyay, et al. (2020), among health workers worldwide, susceptibility to infections is higher among women (71.6%) and nurses (38.6%). However, deaths are mainly in men (70.8%) and doctors (51.4%).

*Table 1.* Profile of COVID-19 Confirmed Cases from Jan – Apr 2020

| Characteristics | Values |
| --- | --- |
| Sex | |
|     Male, n(%) | 3427 (46%) |
|     Female, n(%) | 4032 (54%) |
| Pregnant Women, n(%) | 47 (0.6%) |
| Age, Median (IQR) | 47 (32 – 46.8) |
|     Children (0 – 12 years old) | 128 (2%) |
|     Teens (13 – 19 years old) | 115 (2%) |
|     Working Class/Adult (20 - 60 years old) | 5303 (71%) |
|     Senior Citizens (≥60 years old) | 1913 (26%) |
| Health Worker, n(%) | 1267 (16.99%) |

**Basic Epidemiologic and Clinical Characteristics**

This section examines the epidemiological and clinical characteristics of early cases of COVID-19 in the Philippines. The results are summarized in Table 2. Compared to demographic characteristics, epidemiologic and clinical characteristics have clear implications for hospital resource utilization and mortality of patients. Epidemiologic characteristics, including the onset of symptoms to testing and diagnosis, are used to determine the number of quarantine days for patient isolation. Furthermore, clinical manifestations such as presenting symptoms during onset and comorbidities are used as criteria to classify patient severity and, thereby, determine appropriate care for patients (CDC COVID-19 Response Team, 2020).

The presenting symptom during the onset of COVID-19 infection is a relevant indicator for patient care, including isolation, hospitalization, and use of hospital resources. The most common presenting signs or symptoms observed upon infection are fever (4009, 54%) and cough (4370, 59%). These symptoms, along with other symptoms (1540, 21%) such as headache and fatigue, are considered mild to moderate. Another notable presenting symptom is the difficulty of breathing (1812, 24%), which is usually the reason for hospitalization and the use of mechanical ventilators. Patients who experience difficulty in breathing (dyspnea and or hypoxia) are considered severe to critical cases. Severe to critical cases of COVID-19 have higher case fatality, and require much more attention (Wu et al., 2020). Higher-case fatality is also found in patients with comorbidities such as cardiovascular disease, diabetes, chronic respiratory disease, history of stroke, cancer, lung, and kidney diseases (CDC COVID-19 Response Team, 2020; Wu et al., 2020; Yang et al., 2020). Thus, patients with any of these comorbidities may be classified as severe to critical. In the early stage of COVID-19 in the Philippines, the most commonly reported comorbidities are hypertension (17%) and diabetes (4.4%).

The time from the onset of symptom to testing and diagnosis are relevant measures to assess COVID-19 management. Immediate testing and early diagnosis leads to timely patient isolation and care. Our results show that the average is three (3) days from the onset of symptoms in terms of testing probable cases. However, despite having a short median time from onset to testing, the onset to diagnosis takes a relatively long time. The average is eight (8) days with an interquartile range of four to 12 days.

*Table 2. Basic Epidemiological and Clinical Characteristics*

| Characteristics | Values |
| --- | --- |
| Presenting Symptoms, n(%) | |
| Fever, | 4009 (54%) |
| Cough, | 4370 (59%) |
| Cold, | 1318 (18%) |
| Sore throat, | 1368 (18%) |
| Difficulty Breathing, | 1812 (24%) |
| Diarrhea, | 376 (5.0%) |
| Other Symptoms, | 1540 (21%) |
| The Onset of Symptom to Testing, Median (IQR) days | 3 (1 – 7) |
| The Onset of Symptom to Diagnosis, Median (IQR) days | 8 (4 – 12) |
| Presence of Comorbidities, n (%) | |
| Hypertension | 1282 (17%) |
| Diabetes | 328 (4.4%) |

**Length of Hospital Stay**

The length of hospitalization is a critical indicator of the use of medical and hospital resources dedicated to COVID-19. The longer the patient stays in the hospital, the more resources are consumed or used. This study shows that the average length of hospital stay is 11 days with an interquartile range of five to 18 days. For patients with critical illness or complications, the length of hospital stay is 18 days with an interquartile range of 14 to 24 days (Chen et al., 2021). Furthermore, we examine the difference in the length of hospitalization between male and female patients (Figure 2 left) and between patients whose nature of work is related and not related to health service (Figure 2 right). The boxplot shows that the median length of hospitalization is significantly higher for male patients than for female patients at a 0.05 level of significance. Male patients, whenever admitted, stay longer in hospitals and use more medical and hospital resources than female patients. Examining the presenting symptoms further reveals that difficulty breathing and comorbidities are higher in the male group (difficulty breathing, 31%; hypertension, 40%) than in the female group (difficulty breathing, 18%; hypertension, 25%). These conditions usually require hospitalization. In addition, patients whose nature of work is health service-related do not require more hospital resources. When the length of the hospital stay of health care workers and those whose work is not related to health service is compared, the result shows no statistical difference (p-value =0.69).

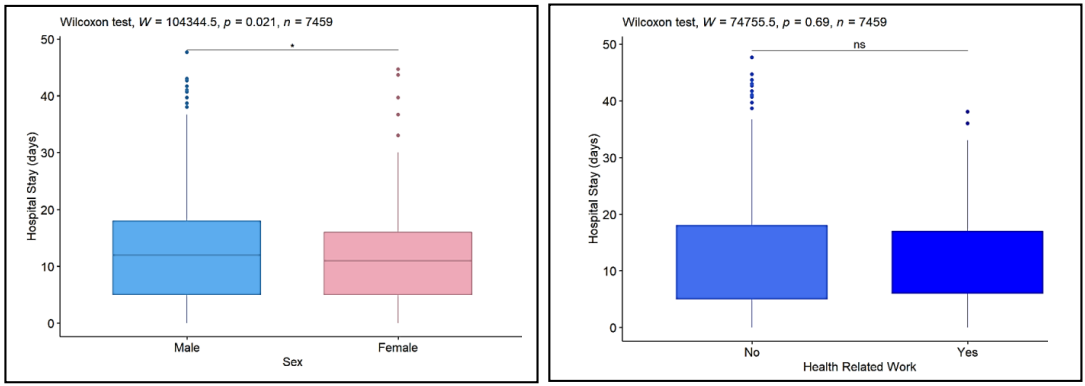

*Figure 2.* Boxplots of the length of hospital stay for Sex (left) and Health-Related Work (right)

Several studies have examined the length of hospitalization according to age groups (Qui et al., 2020; C. Shi et al., 2020; Y. Shi et al., 2020; Wei et al., 2020; Xia et al., 2020). There appears to be little difference in the length of hospital stay between younger and older populations in all of these studies. Our result of this study coincides with the findings of the previous studies. The length of hospitalization in the Philippines during the early stage of COVID-19 is not statistically different across age groups (p-value, 0.09) which implies that utilization of medical and hospital resources across age groups is the same.

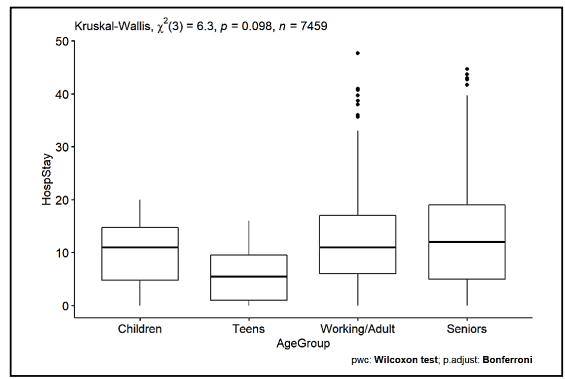

*Figure 3.* Boxplots of the length of hospital stay for different age groups

**Mortality and Survivability**

The status of infected cases is monitored closely to account for the mortality in each segment of the population. In the data that the researcher obtained, the COVID-19 confirmed cases have four categories: Dead, Recovered, Currently Confined, and For Validation. The cases classified For Validation are those cases that need to be checked and verified to ensure if cases have already recovered. The death toll from January to April 2020 has reached 568 people (7.6%) with a median hospital day of around seven days while confirmed recovery rate is 1042 (13.9%) with median hospital days of around 14 days.

Mortality is found to be significantly higher for males (9.2%) compared to females (5.7%). Two statistical reports by the World Bank (2018) explains this. First, the general life expectancy of Filipinos is shorter for males (67.2 years) than for females (75.5 years). Second, the overall mortality rate of Filipinos is higher for males (233.697 per 1,000 males) than for females (129.01 per 1000 females). Several studies have confirmed the association of male sex to COVID-19 deaths (Peckham et al., 2020; Yanez et al., 2020).

Further, several studies have found an association between older age and mortality (Ho et al., 2020; Kang & Jung, 2020; Yanez et al., 2020). The findings of this investigation are similar to these studies. Most of those who died are senior citizens (19.2%). The large value of the chi-square residual for the senior citizens' mortality further indicates higher mortality is associated with this group. Higher mortality is also found to be associated with mild respiratory diseases (Masetti et al., 2020), cancer (Assaad et al., 2020), diabetes, hypertension, and cardiovascular diseases (de Almeida-Pititto et al., 2020). Our analysis reveals that difficulty breathing and diarrhea at the onset of infection and the presence of diabetes and hypertension are risk factors associated with COVID-19 mortality.

*Table 3.* Comparison of COVID-19 outcome among Patient Characteristics

| Characteristics | COVID-19 Outcome | | p-value |
|---|---|---|---|
| | Dead | Survive | |
| Sex, | | | <0.01 |
|     Male, n (%) | 371 (9.2%) | 3661 (90.8%) | |
|     Female, n (%) | 197 (5.7%) | 3230 (94.3%) | |
| Age, n (%) | | | <0.01 |
|     Children (0 – 12 years old) | 3 (2.3%) | 125 (97.6%) | |
|     Teens (13 – 19 years old) | 3 (2.6%) | 112 (97.3%) | |
|     Working Class/Adult (20 - 60 years old) | 194 (3.6%) | 5109 (96.3%) | |
|     Senior Citizens (≥60 years old) | 368 (19.2%) | 1545 (80.8%) | |
| Presenting Symptoms, n (%) | | | |
|     Fever | 369 (9.2%) | 3640 (90.8%) | 0.755 |
|     Cough | 451 (9.5%) | 4279 (90.5%) | 0.094 |
|     Cold | 80 (6.1%) | 1238 (93.9%) | 0.327 |
|     Sore Throat | 60 (4.4%) | 1308 (95.6%) | 0.142 |
|     Difficulty Breathing | 400 (22.1%) | 1412 (77.9%) | <0.01 |
|     Diarrhea | 29 (7.7%) | 347 (92.3%) | 0.045 |
| Presence of Comorbidities, n (%) | | | |
|     Hypertension | 205 (15.9%) | 1077 (83.9%) | <0.01 |
|     Diabetes | 55 (16.7%) | 273 (83.2%) | <0.01 |

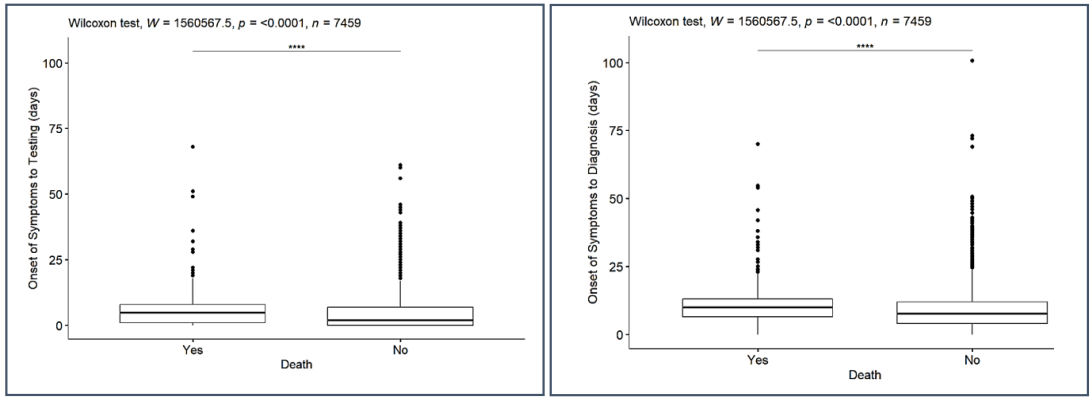

*Figure 4.* Boxplots of onset of symptom to specimen collection and death (left); onset of symptom to diagnosis and death (right)

Further analysis reveals that there is a significant delay in the testing (Figure 4 left) and diagnosis(Figure 4 right) of those who died compared to those who survive (p-values are both <0.001). The average onset to testing days for those who died is five days while those who survived is only two days. The average is ten days for the mortality group for the onset to diagnosis while only seven days for those who survived. The results imply that timely testing and diagnosis may aid in immediate isolation, care, and treatment of COVID-19 patients and prevent some occurrences of mortality among the population. The study of Vecino-Ortiz, et al. (2021) has found that effective contact tracing, testing, and diagnosis will reduce around 10 percent mortality in the population.

Lastly, we determine the risk factors associated with COVID-19 mortality. The result of the survival analysis revealed that older age, male sex, difficulty breathing, having comorbidities like hypertension and diabetes are significant risk factors to higher COVID-19 mortality. There is an increased risk of COVID-19 mortality for older patients. Specifically, there is an equivalent 4.6% increase in the expected hazard for every one-year increase in age. A similar result was obtained in the study of Grasselli, et al. (2020), where older age has a hazard ratio of more than one (1) (HR, 1.75; 95%CI, 1.6 – 1.92).

Furthermore, hypertension and diabetes are two of the most common diseases of Filipinos. The associated risk of these diseases to COVID-19 mortality is examined. Table 4 shows that the expected hazard is 2.311 higher for patients with hypertension, holding other variables constant. In the same way, diabetes also increases the risk of mortality of COVID-19 cases. The expected hazard is 2.694 higher for diabetic patients, holding other factors constant. The result is consistent with the findings of Grasselli, et al. (2020), where type 2 diabetes is also found to be highly associated with mortality (HR, 1.18; 95% CI, 1.01 – 1.39) and with the findings of de Almeida-Pititto, et al. (2020) where hypertension and diabetes are found to be significant risk factors to COVID-19 mortality.

*Table 4.* Survival Analysis of COVID-19 Patients

|  | Hazard Ratio | 95% CI | p-value |
|---|---|---|---|
| Age (≥60years) | 1.051 | 0.915 – 1.1851 | <0.01 |
| Male (Sex) | 1.046 | 1.0388 – 1.0572 | <0.01 |
| Difficulty Breathing (Yes) | 2.767 | 2.1367 – 3.5826 | <0.01 |
| Hypertension (Yes) | 2.311 | 1.0763 – 2.7264 | 0.023 |
| Diabetes (Yes) | 2.694 | 1.3334 – 4.0056 | 0.002 |

## 4.0 Discussion

The vaccination program against COVID-19 aims to prevent mortality, preserve the health system's capacity, and protect the populations most at risk. It is therefore imperative to achieve these goals such that vaccination priority be given to the most susceptible population subgroups with longer hospitalization tendency, and at higher risk of mortality. This study aims to determine the demographic, epidemiologic, and clinical characteristics of COVID-19 cases that are highly susceptible to COVID-19 infection, with longer hospitalization and at higher risk of mortality and to provide insights that may be used to assess the vaccination priority program and allocation of medical and hospital resources.

Results reveal that women are more susceptible than men in the Philippines during the early months of COVID-19. The result is similar to the data in several European countries having higher rates of COVID-19 infection among women. Denmark, Portugal, Spain, and Switzerland have recorded 54%, 57%, 51%, and 53%, respectively (Sandoiu, 2020). However, virologists and epidemiologists believed that women have an apparent advantage over men regarding susceptibility to COVID-19. The explanation ranges from biological and genetic factors to epidemiological and behavioral ones (Wenham et al., 2020). In contrast, Bertocchi (2020) proposed an explanation as to why some countries, including the Philippines, have recorded higher infection rates among women than men. She argued that the socio-economic profile and the employment rate can explain the difference in susceptibility between men and women. In countries where the employment rate gap between men and women is small, the number of exposed women in the workplace would also be closed to the number of exposed men. Thus, given the same amount of exposure to the virus, women are more susceptible than men, primarily when higher transmission of COVID-19 infections occurs in workplaces. Although women are more susceptible to the virus, men have longer hospitalization days and are at higher risk of mortality. An explanation for this is found by examining the presenting symptoms associated with severe COVID-19 infection that requires hospitalization. Presenting symptoms and conditions such as difficulty breathing and comorbidities are higher in the male group. The higher prevalence of smoking among men in the population, 29.7% compared to only 8.1% among women, may also offer a viable explanation for their prolonged hospital stay and higher mortality. In addition, men have a shorter life expectancy compared to women in the Philippines.

On the other hand, the susceptibility of the population according to age groups is a much-needed discussion because age is included as a criterion for vaccination priority. In the prioritization framework released by DOH, senior citizens (≥60 years old) are listed in the second topmost priority (A2) while the working class, aged 20 to 60 years old, despite being the group with the highest rate of infection is not in the priority list unless of course, they qualify in the other criteria for vaccination priority such as having comorbidity. Senior citizens are considered vulnerable as they are prone to have comorbidities associated with the severe outcome of COVID-19 infection. However, senior citizens have minimal mobility thus, with limited exposures to COVID-19 as they cannot go out except to buy food and medicine. On the other hand, the working class, specifically those whose nature of work requires them to move around (service sector jobs) constantly, have greater susceptibility and may be at a higher risk of being infected. Although the working class, in general, will most likely survive COVID-19 and will not likely have a severe outcome, the quarantine time of at least 14 days also has a significant effect to the economy in the long run. Most of those who work in the service sector are on a no-work-no-pay basis which constitutes to a significant fraction of around 58.1 percent of the population (low-income group) as estimated by Albert, et al. (2018). Furthermore, senior citizens have a higher risk of COVID-19 mortality compared to other age groups which is another reason why this group is placed next to the front line health workers in the vaccination priority.

Moreover, frontline health workers are placed at the top priority (A1) of the COVID-19 vaccination program. They are considered highly susceptible due to the nature of their work. The decision to place the frontline health workers on top priority is considered a wise measure in preserving health system capacity. When people protect the health workers, they also protect the most critical essential services. People with comorbidity are placed third (A3) in the vaccination priority. Our analysis further reveals that comorbidities like diabetes and hypertension are risk factors for COVID-19 mortality. These are two of the most common diseases among Filipinos. Hypertension landed in the top leading causes of mortality in the Philippines for 2019. The government exerts efforts to reduce hypertension mortality such as free maintenance medicine and other health care services. With COVID-19 infection, a patient with hypertension and diabetes are given priority to medical care and hospitalization to reduce risk of mortality. The high association between comorbidity and mortality also explains why people with comorbidities are placed in priority A3 in the vaccination program released by the DOH. Protecting these segments of the population also saves depletion in hospital resources in case of a surge.

Lastly, management and planning of COVID-19 patients and facilities are affected by several factors including early detection and diagnosis of COVID-19. Immediate testing leads to an early diagnosis and to a timely patient isolation and provision of medical care. Results of this study reveal that there is a delay in the testing and diagnosis of COVID-19 patients who died. Wei, et al. (2020) argued that there is a correlation between COVID-19 mortality and testing covering which implies that mortality may be reduced by increasing testing coverage. Several factors may explain the delay in the diagnosis. This includes the limited available facilities that can carry out laboratory analysis of the specimen during the early stage of COVID-19 in the country. DOH has then increased the number of COVID-19 testing laboratories and centers to minimize delays in the testing and detection of COVID-19. As of this writing, there are around 256 accredited testing centers (Department of Health, 2021).

**5.0 Conclusions and Recommendation**

In situations where the amount of exposure between men and women due to employment is close to equal, women are more susceptible to COVID-19 infection. In addition, the working or adult group is the most susceptible age group due to higher exposures in the workplaces and frequent mobility in the community. Therefore, because those with greater exposure are more susceptible to COVID-19 infection, it is recommended that susceptibility due to exposure is considered or included in the criteria of determining priority subgroups for vaccination against COVID-19 and other potentially deadly viruses in the future.

In cases where both men and women are hospitalized, men require and consume more medical resources compared to women. The length of hospital stay is just one of the many measures of medical resource utilization. Other necessary measures are the ICU facilities and use of specific resources such as ventilators and laboratory facilities. Another study may be carried out assessing the cost of the use of these resources. It is further recommended that medical resources be utilized in determining the priority subgroups for vaccination.

Risk factors associated with increased mortality are older age, male sex, hypertension and diabetes, and difficulty breathing. It is recommended that patients with these combined demographic, comorbidities, and presenting symptoms be given priority in hospitalization and admission to the COVID-9 facility for proper care and monitoring. Lastly, immediate testing and early diagnosis are crucial to the survival of COVID-19 patients. Therefore, efficient contact tracing for isolation, testing, and diagnosis must be a top priority in the COVID-19 response strategy.

**Acknowledgment**

The researcher would like to acknowledge the following: a) the Research Division of the University of Southeastern Philippines for the financial support to carry out this research project in 2020, and b) the DOST-PCIEERD for the training grant Learning at Scale Vol. 1: Data Science Track Extension in 2018. The training has provided the researcher the statistical computing skills to carry out the data analysis in this project.

**Appendix 1**

| Order of Priority | Population Groups |
|---|---|
| A1 | Frontline workers in health facilities both national and local, private and public, health professionals and non-professionals like students, nursing aides, janitors, barangay health workers, etc. |
| A2 | Senior citizens aged 60 years old and above. |
| A3 | Persons with comorbidities not otherwise included in the preceding categories |
| A4 | Frontline personnel in essential sectors including uniformed personnel and those in working sectors identified by the IATF as essential during ECQ |
| A5 | Indigent population not otherwise included in the preceding categories |
| B1 | Teachers, social workers |
| B2 | Other government workers |
| B3 | Other essential workers |
| B4 | Socio-demographic groups at significantly higher risk other than senior citizens and indigent people |
| B5 | Overseas Filipino Workers |
| B6 | Other remaining workforce |
| B7 | Rest of the Filipino population not otherwise included in the above groups |